\documentclass[twocolumn,aps,prl,showpacs]{revtex4}

\usepackage{graphicx}
\usepackage{graphics}
\usepackage{epsfig}
\usepackage[latin1]{inputenc}
\usepackage{latexsym}
\usepackage{amsmath}
\usepackage{amssymb}
\usepackage{enumerate}

\begin{document}
\title{Proposed Experiment for Testing Quantum Contextuality with Neutrons}
\author{Ad\'{a}n Cabello}
\email{adan@us.es}
\affiliation{Departamento de F\'{\i}sica Aplicada II,
Universidad de Sevilla, E-41012 Sevilla, Spain}
\author{Stefan Filipp}
\email{sfilipp@ati.ac.at}
\affiliation{Atominstitut der {\"{O}}sterreichischen Universit{\"{a}}ten,
Stadionallee 2, A-1020 Vienna, Austria}
\author{Yuji Hasegawa}
\email{hasegawa@ati.ac.at}
\affiliation{Atominstitut der
{\"{O}}sterreichischen Universit{\"{a}}ten, Stadionallee 2, A-1020
Vienna, Austria}
\affiliation{PRESTO, Japan Science and Technology
Agency (JST), Kawaguchi, Saitama 332-0012, Japan}
\author{Helmut Rauch}
\email{rauch@ati.ac.at}
\affiliation{Atominstitut der
{\"{O}}sterreichischen Universit{\"{a}}ten, Stadionallee 2, A-1020
Vienna, Austria}

\date{\today}


\begin{abstract}
We show that an experimental demonstration of quantum contextuality
using 2 degrees of freedom of single neutrons based on a violation
of an inequality derived from the Peres-Mermin proof of the
Kochen-Specker theorem would be more conclusive than those obtained
from previous experiments involving pairs of ions [M. A. Rowe {\em
et al.}, Nature (London) {\bf 409}, 791 (2001)] and single neutrons
[Y. Hasegawa {\em et al.}, Nature (London) {\bf 425}, 45 (2003)]
based on violations of Clauser-Horne-Shimony-Holt-like inequalities.
\end{abstract}



\maketitle


Bell inequalities are constraints imposed by local hidden-variable
theories (LHVTs) on the values of some specific linear combinations
of the averages of the results of spacelike separated experiments on
distant systems. From a conclusive experimental violation of a Bell
inequality we would learn that no local hidden-variable theory can
describe the world we see. This ``quantum nonlocality'' (i.e., the
fact that quantum mechanics [QM] cannot be reproduced by LHVTs
\cite{Bell64}) is supposed to be universal, in the sense that we
should be able to detect it by using {\em any} composite quantum
system (photons, electrons, neutrons, atoms, molecules). However, in
practice, so far we do not have a loophole-free experimental
violation of a Bell inequality, and, specifically we do not have any
Bell inequality test involving spacelike separated experiments with
any quantum system apart of photons \cite{ADR82,TBZG98,WJSWZ98}.
Therefore, an interesting question is what can we learn about the
possibility of reproducing experimental observations with
hidden-variable theories from experiments performed on massive
nonspacelike separated quantum systems such as neutrons and atoms.

LHVTs are a subset of a larger class of hidden-variable theories
known as noncontextual hidden-variable theories (NCHVTs). NCHVTs are
defined as those in which the result of a measurement of an
observable is assumed to be predetermined and not affected by a
(previous or simultaneous) suitable measurement of any other {\em
compatible} (i.e., comeasurable) observable.

In this sense, some recent experiments testing Bell inequalities
on pairs of two-level $^9$Be$^+$ ions separated by a distance of 3
$\mu$m \cite{RKMSIMW01} and two two-level degrees of freedom
(spatial and spin components) of single neutrons \cite{HLBBR03}
cannot be considered tests of LHVTs, but only tests of NCHVTs
(see, for instance \cite{Vaidman01,Santos05}). Both mentioned
experiments have in common the fact that they test the
Clauser-Horne-Shimony-Holt Bell inequality \cite{CHSH69},
\begin{equation}
|\langle A_0 B_0 \rangle+\langle A_0 B_1 \rangle+\langle A_1
B_0\rangle-\langle A_1 B_1 \rangle| \le 2.
\end{equation}
Common to both experiments is also the fact that they have very good
overall detection efficiency, both around $99\%$ (specially when
compared with that of Bell tests with photons which is around $5\%$
\cite{WJSWZ98}). This very good detection efficiency would allow us
to avoid the fair sampling assumption (i.e., that the detected
particles are a representative sample of all emitted ones) needed to
reach any conclusion in photon experiments of LHVTs
\cite{ADR82,TBZG98,WJSWZ98}. However, common to both experiments is
also the fact that, due to experimental imperfections (imperfect
state creation, manipulations, and detection), the final
experimental results ($2.25 \pm 0.03$ \cite{RKMSIMW01} and $2.051
\pm 0.019$ \cite{HLBBR03}) are indeed closer to the hidden-variable
bound, $2$, than to the maximal violation predicted by QM, $2
\sqrt{2} \approx 2.82$. Therefore, an interesting question is
whether we can design a better ``quantum contextuality'' experiment
(i.e., showing us that no NCHVT can describe the world we see) for
these systems. By a better experiment we mean one in which the
observed violation expected from the hidden-variable bound would be
significantly higher than those obtained in these previous
experiments.

One possible answer comes from the fact that the observation that QM
cannot be reproduced by NCHVTs has its own history starting from
Kochen-Specker theorem \cite{Specker60,Bell66,KS67} and has
developed its own proofs apart from those of Bell's theorem. These
proofs are specifically designed to stress the difference between QM
and NCHVTs. In this Letter, we show that it is possible to
demonstrate quantum contextuality using massive quantum systems, and
specifically 2 degrees of freedom of single neutrons following the
steps of a state-dependent multiplicative proof of the KS theorem
\cite{CG96} proposed by Peres \cite{Peres90} and Mermin (who
extended Peres' proof into a state-independent proof
\cite{Mermin90,Mermin93}, which can be converted into a standard
proof of the KS theorem \cite{Peres91}, which indeed contains the
simplest possible \cite{CPP07} standard proof of the KS theorem
\cite{CEG96}).

The Peres-Mermin proof, suitably adapted, for instance, to the case
of the two two-level degrees of freedom (spatial and spin
components) of single neutrons \cite{HLBBR03} is as follows.
Consider a system prepared in the state
\begin{equation}
|\Psi \rangle = \frac{1}{\sqrt{2}}\left(|\downarrow\rangle \otimes
|I\rangle - |\uparrow\rangle \otimes |II\rangle \right),
\label{state}
\end{equation}
where $|\uparrow\rangle$ and $|\downarrow\rangle$ denote the up-spin
and down-spin states, and $|I\rangle$ and $|II\rangle$ denote the
two beam paths in an interferometer. $s$ stands for spin and $p$ for
path. Let us suppose that the {\em six} observables $\sigma_x^s$,
$\sigma_y^s$, $\sigma_x^p$, $\sigma_y^p$, $\sigma_x^s \sigma_y^p$,
and $\sigma_y^s \sigma_x^p$ have predefined values $v(\sigma_x^s)$,
$v(\sigma_y^s)$, $v(\sigma_x^p)$, $v(\sigma_y^p)$, $v(\sigma_x^s
\sigma_y^p)$, and $v(\sigma_y^s \sigma_x^p)$, respectively. For the
state $|\Psi \rangle$, the predictions of QM used in the proof are
the following five:
\begin{subequations}
\begin{align}
v(\sigma_x^s) v(\sigma_x^p) & = -1,
\label{uno}\\
v(\sigma_y^s) v(\sigma_y^p) & = -1,
\label{dos}\\
v(\sigma_x^s \sigma_y^p) v(\sigma_x^s) v(\sigma_y^p) & = 1,
\label{tres}\\
v(\sigma_y^s \sigma_x^p) v(\sigma_y^s) v(\sigma_x^p) & = 1,
\label{cuatro}\\
v(\sigma_x^s \sigma_y^p) v(\sigma_y^s \sigma_x^p) & = -1.
\label{cinco}
\end{align}
\end{subequations}
Equations (\ref{uno}) and (\ref{dos}) follow from the
anticorrelations designed into the state $|\Psi \rangle$, Eqs.
(\ref{tres}) and (\ref{cuatro}) follow from the fact that the values
assigned to mutually commuting operators whose product is the
identity must obey the same relation satisfied by the operators
themselves, and Eq. (\ref{cinco}) follows from the fact that
$(\sigma_x^s \otimes \sigma_y^p) (\sigma_y^s \otimes
\sigma_x^p)=\sigma_z^s \otimes \sigma_z^p$ and the anticorrelations
of the state $|\Psi \rangle$.

To show that it is impossible to ascribe predefined values $-1$ or
$1$ to each and every of these six observables, it is enough to
multiply both sides of Eqs. (\ref{uno})--(\ref{cinco}): Each
observable appears twice; therefore, we have $1$ as the product of
the left-hand sides. However, we have $-1$ as the product of the
right-hand sides \cite{Peres90,Mermin90,Mermin93}.

An ideal experiment for discarding NCHVTs would be simply to confirm
each and every of these {\em five} predictions of QM. An important
point is that each equation corresponds to a different experimental
context. Therefore, the experiment requires to perform five
different types of experiments. A fundamental point is the
measurement apparatus used for measuring, e.g., $\sigma_x^s
\sigma_y^p$ {\em must be the same} in the experimental context
corresponding to Eq. (\ref{tres}) and in that corresponding to Eq.
(\ref{cinco}).

Then, since the perfect correlations (or anticorrelations) on which
the proof is based are not obtained in real experiments, one should
derive a Bell-like inequality. Some inequalities for quantum
contextuality have been proposed
\cite{BBKH01,SBZ01,Larsson02,Cereceda02}. The one introduced here
has two advantages: it is the direct translation of the Peres-Mermin
proof into an experimentally testable inequality, and provides a
significant contrast between the prediction of QM and the
hidden-variable bound.

The relevant properties of the Peres-Mermin proof come from Eqs.
(\ref{uno})--(\ref{cinco}). Then, a reasonable choice is to
investigate the inequality obtained by the linear combination of the
averages of the results obtained in the five experimental contexts,
with the same coefficients ($-1$ or $1$) appearing in the quantum
predictions [i.e., in Eqs. (\ref{uno})--(\ref{cinco})]. For this
inequality, it can be proven (for instance, by writing a simple
computer program) that, in any NCHVT,
\begin{eqnarray}
-\langle\sigma_x^s \cdot \sigma_x^p\rangle
-\langle\sigma_y^s \cdot \sigma_y^p\rangle
+\langle\sigma_x^s \sigma_y^p \cdot \sigma_x^s \cdot \sigma_y^p \rangle & & \nonumber \\
+\langle\sigma_y^s \sigma_x^p \cdot \sigma_y^s \cdot \sigma_x^p
\rangle -\langle\sigma_x^s \sigma_y^p \cdot \sigma_y^s
\sigma_x^p\rangle & \le & 3, \label{ineq}
\end{eqnarray}
while the prediction of QM is just $5$, as can be seen from the
predictions of QM leading to Eqs. (\ref{uno})--(\ref{cinco}). This
is the inequality that one should test. Such a test requires five
different types of experiments and requires measuring any observable
using the same apparatus independently of the context in which it
appears.

However, it is interesting to note that the five experiments in Eqs.
(\ref{uno})--(\ref{cinco}) play different roles in the proof. Note
the difference between Eqs. (\ref{tres}) and (\ref{cuatro}) and the
other three equations (\ref{uno}), (\ref{dos}), and (\ref{cinco}).
While Eqs. (\ref{tres}) and (\ref{cuatro}) should hold for {\em any}
preparation (i.e., are state-independent predictions of both QM and
NCHVTs), Eqs. (\ref{uno}), (\ref{dos}), and (\ref{cinco}) are
predictions specific for a particular quantum state (\ref{state}).

Since Eqs. (\ref{tres}) and (\ref{cuatro}) hold both in QM and in
NCHVTs and do not depend on any particular preparation of the
state, we do not really need to test them. This observation was
first made in \cite{CG98} and was used in several Bell-like
inequalities for NCHVTs. Therefore, it is enough to test the
following Bell-like inequality:
\begin{equation}
-\langle\sigma_x^s \cdot \sigma_x^p\rangle -\langle\sigma_y^s \cdot
\sigma_y^p\rangle -\langle\sigma_x^s \sigma_y^p \cdot \sigma_y^s
\sigma_x^p\rangle \le 1. \label{ineq2}
\end{equation}
However, and this is an important point missed in previous
discussions, an experiment of this type cannot be considered a test
of quantum contextuality unless we also describe how, at least in
principle, the two predictions not considered in the Bell-like
inequality (\ref{ineq2}) can be tested using the same measuring
apparatus used to measure the observables in the inequality
(\ref{ineq2}). This means that, for instance, the measuring
apparatus used for measuring $\sigma_x^s \sigma_y^p$ must be the
same regardless of whether it is measured together with $\sigma_x^s$
and $\sigma_y^p$, or together with $\sigma_y^s \sigma_x^p$. Without
such a prescription, an experiment to test the inequality
(\ref{ineq2}) cannot be considered a contextuality experiment, since
the six observables involved are just tested in one context and no
description of how they can be tested in a different context is
provided.

Another motivation for this Letter comes from the fact that a
previous experiment on quantum contextuality with single neutrons
\cite{HLBBR06} does not satisfy the above requirements.
Specifically, in \cite{HLBBR06} the measurement of $\sigma_x^s
\sigma_y^p \cdot \sigma_y^s \sigma_x^p$ was accomplished by
measuring $\sigma_z^s \sigma_z^p$. This method is unsuitable,
because a measurement of $\sigma_z^s \sigma_z^p$ gives only the
value of the product $\sigma_x^s \sigma_y^p \cdot \sigma_y^s
\sigma_x^p$, and because the observable $\sigma_z^s \sigma_z^p$ is
not compatible with the observables $\sigma_x^s$ and $\sigma_y^p$,
and therefore $\sigma_z^s \sigma_z^p$ cannot be used (even
potentially) to perform the subsequent required measurements (e.g.,
$\sigma_x^s \sigma_y^p$ together with $\sigma_x^s$ and
$\sigma_y^p$).

Measuring $\sigma_x^s \sigma_y^p$ and $\sigma_y^s \sigma_x^p$, the
two observables in Eq. (\ref{cinco}), simultaneously is typically
not a problem (see, e.g., \cite{CBPMD05,YZZYZZCP05}). The challenge
is to prove that the apparatuses used for measuring one of them can
be combined (potentially) with the subsequent measurements required
in Eqs. (\ref{tres}) and (\ref{cuatro}) (e.g., to show that the
apparatus for measuring $\sigma_x^s \sigma_y^p$ can be combined with
a subsequent measurement of $\sigma_x^s$ and $\sigma_y^p$).


\begin{figure}[t!!]
\centerline{\includegraphics[width=\columnwidth]{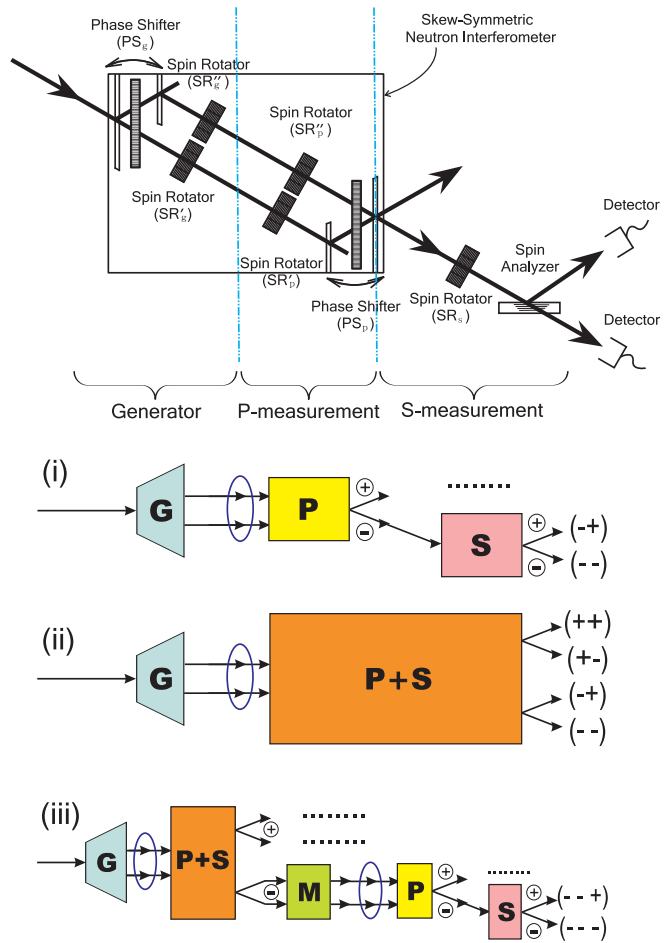}}
\caption{\label{Fig1} Above: A proposed experimental setup with a
neutron interferometer. The interferometer has two functions: the
first half as a state generator, and the second half as a path
measurement apparatus. In both parts, a phase shifter (PS) as well
as a pair of spin rotators (SR) which enable relative phase and
spinor manipulations are inserted. The outgoing beam suffers a spin
measurement. Below: Three diagrams for the different measurements
required to test the inequality (\ref{ineq}). (i) For measurements
of $\sigma_x^s \cdot \sigma_x^p$ and $\sigma_y^s \cdot \sigma_y^p$:
After going through a state generator ($G$), a state suffers a path
measurement ($P$) followed by a spin measurement ($S$).
Consequently, each outgoing beam gives the results of the two
measurements. (ii) For measurements of $\sigma_y^s
\sigma_x^p\cdot\sigma_x^s \sigma_y^p$: By tuning one of the spin
rotators to a spin-flip operation in the path measurement part, the
second half of the interferometer together with a spin analyzer
(P+S) work as a Bell-state discriminator: four outgoing beams are
assigned to the four possible results of the measurements. (iii) For
measurements of $\sigma_y^s \sigma_x^p \cdot \sigma_y^s \cdot
\sigma_x^p$ and $\sigma_x^s \sigma_y^p \cdot \sigma_x^s \cdot
\sigma_y^p$: After the apparatus $P+S$, a state mixer ($M$)
eliminates the former information about the result of either
observable, and is followed by a path and a spin measurement.}
\end{figure}


In the following we will describe how to test the inequality
(\ref{ineq2}) using two two-level degrees of freedom (spatial and
spin components) of single neutrons, and satisfying all the above
mentioned requirements for a proper NCHVT interference experiment.
(General descriptions of neutron interferometer experiments are
summarized in the literature \cite{RW00}.) A schematic drawing,
together with diagrams of the experiments is depicted in Fig.
\ref{Fig1}. The first half of the interferometer is used for a
(Bell) state preparation and the second half is used for the
measurement of the path observable, e.g., $\sigma_{x}^p$. In
addition to auxiliary phase shifters, a pair of spin rotators
(dc-coils), which enables arbitrary relative phase and spinor
manipulation, is inserted in both parts. After the path measurement,
the spinor component, e.g., $\sigma_{x}^s$, is measured by the use
of a conventional spin analysis system, namely, a spin analyzer
accompanied by a spin rotator. Combinations of these path and spin
measurements will accomplish the measurements of $\sigma_x^s \cdot
\sigma_x^p$ and $\sigma_y^s \cdot \sigma_y^p$ [illustrated in
diagram (i)]. When a spin rotator in one of the beams in the
interferometer is tuned to a spin-flip operation and a suitable spin
analysis is activated, one can perform the measurement of
$\sigma_x^s \sigma_y^p \cdot \sigma_y^s \sigma_x^p$
\cite{CBPMD05,YZZYZZCP05} [cf. diagram (ii)]. The four outcomes
correspond to a full Bell-state analysis.

Since it is essential to state a prescription of how one can realize
the measurements corresponding to Eqs. (\ref{tres}) and
(\ref{cuatro}) with the same apparatus as that corresponding to Eq.
(\ref{cinco}), we describe a possible setup for a neutron
interferometer experiment. Our proposal consists of using exactly
the same apparatus used in case (ii) for measuring $\sigma_y^s
\sigma_x^p\cdot\sigma_x^s \sigma_y^p$, followed by a path and spin
measurements after going through a ``state-mixer'' [shown in case
(iii)]. In practice, the ``state-mixer'' is a beam splitter plus an
unitary state rotation. This device forms a quantum eraser
\cite{SEW91}, which eliminates the information about the results of
the noncommuting observable, i.e., $\sigma_x^s \sigma_y^p$ in case
one would want to measure $\sigma_y^s
\sigma_x^p\cdot\sigma_y^s\cdot\sigma_x^p$ and vice versa: this, in
turn, enables measurement of other comeasurable observables. The
function of this component is given by a projection operator to a
state, e.g., $|\Psi' \rangle = |\Psi_1 \rangle +|\Psi_2 \rangle$,
where $|\Psi_{1,2} \rangle$ are eigenstates of the joint measurement
of $\sigma_x^s \sigma_y^p$ and $\sigma_y^s \sigma_x^p$. It is worth
noting here that the device $P+S$ together with the mixer $M$
practically works as a $\sigma_x^s \sigma_y^p$ or $\sigma_y^s
\sigma_x^p$ measurement apparatus. This can be tested easily in a
separate experiment. As a whole, scheme (iii) can be viewed as the
$\sigma_x^s \sigma_y^p$ (or $\sigma_y^s \sigma_x^p$) measurement
followed by the $\sigma_x^s$ and $\sigma_y^p$ (or $\sigma_y^s$ and
$\sigma_x^p$) measurements. Moreover, the combined measurement
scheme (ii) can, in principle, be replaced by these two $\sigma_x^s
\sigma_y^p$ and $\sigma_y^s \sigma_x^p$ measurements in series.

Using data of previous experiments \cite{HLBBR06} for the first two
averages in (\ref{ineq2}) and estimating the result for the third
average from previous experiments on Bell-state discrimination, we
estimate that we will find an experimental value above $2.1$ (vs. a
bound of $1$ for NCHVTs), clearly proving quantum contextuality as
nowhere else before. The expected value would provide an even
clearer quantum contextuality than a recent Mermin-like experiment
involving 3 degrees of freedom \cite{LKSFBRH07}.

Summing up, the contributions of this Letter are the following: We
have pointed out that there are better tools to test NCHVTs than the
CHSH Bell inequality used in recent experiments with pairs of ions
or single photons. A particularly well suited tool is the
Peres-Mermin proof \cite{Peres90,Mermin90,Mermin93}. We have
described how to implement all the steps required for the
Peres-Mermin proof using 2 degrees of freedom of single neutrons.
Specifically, we have provided an explicit description of a
procedure for measuring observables like $\sigma_x^s \sigma_y^p$,
which can be followed by a subsequent measurement of the compatible
observable $\sigma_y^s \sigma_x^p$ or by a subsequent measurement of
the compatible observables $\sigma_x^s$ and $\sigma_y^p$. In
addition, we have derived an inequality which contains the essence
of the Peres-Mermin proof, and can be applied to real experiments.
All these pieces together result in a proposal of an experiment to
test NCHVTs with single neutrons which would presumably improve the
results obtained in previous tests with massive quantum systems.


The authors thank K. Svozil for discussions. This research has been
supported partly by the Austrian Science Foundation (FWF), Projects
No. F1513, No. P18943-N20, and No. P17803-N02. A.C. acknowledges
support from the Spanish MEC Project No. FIS2005-07689, and the
Junta de Andaluc\'{\i}a Excellence Project No. P06-FQM-02243. Y.H.
acknowledges support from the Japanese Science and Technology Agency
(JST).



\end{document}